\documentclass[conference]{IEEEtran}
\IEEEoverridecommandlockouts
\usepackage{cite}
\usepackage{amsmath,amssymb,amsfonts}
\usepackage{graphicx}
\usepackage{textcomp}
\usepackage{xcolor}
\usepackage{flushend}
\usepackage{balance}
\usepackage{epsfig}
\usepackage{epstopdf}
\usepackage{comment}
\usepackage[linesnumbered,ruled,vlined]{algorithm2e}
\usepackage{float}
\usepackage{subfig}
\usepackage{times}
\usepackage{url}

\usepackage{bm}
\usepackage{dsfont} 

\usepackage{tikz}
\usetikzlibrary{arrows.meta,positioning,fit,calc}

\def\BibTeX{{\rm B\kern-.05em{\sc i\kern-.025em b}\kern-.08em
    T\kern-.1667em\lower.7ex\hbox{E}\kern-.125emX}}

\begin{document}

\title{Spatial-Temporal Learning-Based Distributed Routing for Dynamic LEO Satellite Networks \vspace{-0.1in} 
\thanks{This work was supported in part by the National Science and Technology Council (NSTC) of Taiwan under Grant 113-2926-I-001-502-G and 114-2221-E-003-033.}
}
\author{\IEEEauthorblockN{Po-Heng Chou$^{1,3}$, Chiapin Wang$^{2}$, Shou-Yu Chen$^{2}$, Hsiang-Ming Wang$^{2}$, and Kuan-Fu Chen$^{2}$}
\IEEEauthorblockA{
$^{1}$Research Center for Information Technology Innovation (CITI), Academia Sinica (AS), Taipei 11529, Taiwan\\
$^{2}$Department of Electrical Engineering, National Taiwan Normal University (NTNU), Taipei 106308, Taiwan\\
$^{3}$Bradley Department of Electrical and Computer Engineering (ECE), Virginia Tech (VT), Alexandria, VA 22305, USA\\
E-mails: d00942015@ntu.edu.tw, chiapin@ntnu.edu.tw, tata14753369@gmail.com,\\ waynew891@gmail.com, smart3684@gmail.com\vspace{-0.2in}
}
}

\maketitle

\begin{abstract}
This paper proposes a spatial-temporal
learning-based routing framework for dynamic
low Earth orbit (LEO) satellite networks.
The routing problem is formulated as a
partially observable Markov decision process
(POMDP), where each satellite independently
selects its next hop using network-state
observations and destination information.
A graph attention network (GAT) extracts
spatial features from the dynamic satellite
topology and queue states, while a long
short-term memory (LSTM) network captures
their temporal evolution. The resulting
representation is integrated into a deep
Q-network (DQN) for destination-aware
next-hop selection without a separate
traffic-prediction stage.
Simulations in a 45-satellite LEO constellation
under normal and bursty traffic demonstrate
packet delivery performance comparable to a graph neural network (GNN)-enhanced routing baseline, although
the proposed method incurs higher delay
and queue occupancy. Ablation studies
further demonstrate the complementary
benefits of spatial and temporal learning.
\end{abstract}


\section{Introduction}

Low Earth Orbit (LEO) satellite networks have emerged as a key enabler for next-generation global communication systems, providing low-latency, wide-area coverage and seamless connectivity for applications such as Internet of Things (IoT), remote sensing, and disaster recovery~\cite{Chen2026HMLC}.
Compared to traditional geostationary systems, LEO constellations benefit from shorter propagation distances and flexible deployment, making them a key networking layer of future artificial intelligence (AI)-native non-terrestrial networks (NTNs) and space--air--ground integrated networks (SAGINs)~\cite{Dong2023DRL}.

Despite these advantages, the highly dynamic topology of LEO satellite networks poses significant challenges for routing design.
Continuous satellite movement and time-varying inter-satellite links (ISLs) cause network connectivity to evolve rapidly, leading to frequent route disruptions and unstable transmission performance.
Conventional routing approaches, such as shortest-path-based or static routing schemes, may not adequately adapt to rapidly evolving topology and traffic conditions, resulting in increased delay, congestion, and packet loss~\cite{Xu2022SpatialLocationAidedFDR}.
Moreover, centralized routing strategies introduce excessive signaling overhead and suffer from scalability limitations in large-scale constellations~\cite{Chen2025TransformerMIXLEO}.

To address these challenges, reinforcement learning (RL)~\cite{Kaelbling1996} has been widely adopted for sequential decision-making in dynamic environments. RL enables agents to learn adaptive decision policies through interaction with the environment, making it suitable for dynamic LEO network optimization. Building upon this paradigm, deep reinforcement learning (DRL)~\cite{Mnih2015DQN} employs deep neural networks, such as a deep Q-network (DQN), to approximate value functions in high-dimensional state spaces. Beyond routing, DQN-based decision frameworks have also been investigated for other LEO network functions, such as adaptive satellite weighting for multi-satellite positioning~\cite{Chou2026ICCWkshps}. Focusing on routing, distributed DQN- and multi-agent reinforcement learning (MARL)-based schemes enable satellites to select next-hop nodes using local observations and reduce the signaling burden associated with centralized control~\cite{Zuo2021DRLLEORouting, Xu2022SpatialLocationAidedFDR, Chen2025TransformerMIXLEO, Li2025ParetoMARLQueueing}. However, many of these methods rely primarily on instantaneous or short-term local observations and provide limited explicit modeling of temporal congestion evolution.

A second line of research employs graph-aware learning to represent the non-Euclidean structure of satellite networks. Graph neural network (GNN)- and graph attention network (GAT)-enhanced routing methods aggregate neighboring node and link states to improve topology-aware decision-making~\cite{Zhang2025GRLR, Ran2025GNNEnhancedMARL, Rao2025DeepGraphAttentionIES}. Temporal graph convolution has also been introduced to capture dynamic topology evolution and accelerate multi-agent routing training~\cite{Xiang2025TemporalGraphConvRouting}. Nevertheless, in several graph-aware routing methods, temporal traffic evolution is modeled indirectly, embedded in topology-state updates, or not integrated into the same next-hop value-learning process used for routing decisions.

A third line of research relies on prediction-assisted routing. Spatiotemporal traffic and network-state prediction models have been developed to estimate future congestion conditions and guide routing decisions~\cite{Ju2025SpatioTemporalTrafficPrediction, Wang2025STStatePredictionRouting}. Delay-aware routing frameworks further combine traffic prediction with distributed control to improve latency and packet delivery performance~\cite{Li2026DelayAwareLEOIoT}. However, prediction and routing are commonly optimized as separate stages. Such separation may introduce a prediction--decision mismatch when the predicted state becomes outdated before routing execution, particularly under rapidly changing topology and bursty traffic conditions.

Despite these advances, existing routing
methods still face challenges in jointly
modeling spatial topology, temporal network
dynamics, and adaptive next-hop decisions.
To address these limitations, we propose
a spatial-temporal learning-based distributed
routing framework that integrates GAT-based
spatial feature extraction, LSTM-based
temporal modeling, and DQN-based
destination-aware next-hop selection without
a separate traffic prediction stage.
Using network-wide observations, each
satellite independently selects a feasible
next hop rather than relying on centralized
end-to-end route computation.
Following prior work on dynamic LEO
routing~\cite{Zuo2021DRLLEORouting},
we formulate the routing problem as a
partially observable Markov decision process
(POMDP) to account for unobserved traffic
dynamics and uncertainty in network conditions.

The main contributions of this paper are
summarized as follows:
\begin{itemize}
    \item \textbf{Unified spatial-temporal
    representation:}
    We integrate GAT-based spatial feature
    extraction with LSTM-based temporal
    modeling to capture dynamic topology
    and congestion evolution in LEO
    satellite networks.

\item \textbf{Distributed decision-making:}
We develop a destination-aware GAT--LSTM--DQN
framework that maps spatial-temporal observations
to feasible next-hop actions. Each satellite
independently performs next-hop selection without
centralized end-to-end path computation.

\item \textbf{Performance evaluation:}
We evaluate the proposed framework in a 45-satellite
LEO network under normal and bursty traffic conditions.
Comparisons with adapted learning-based baselines
demonstrate competitive packet delivery performance,
while ablation studies validate the benefits of
combining spatial and temporal learning.

\end{itemize}

\section{System Model}
\label{sec:system_model}

\begin{figure}[t]
\centering
\includegraphics[width=0.48\textwidth]{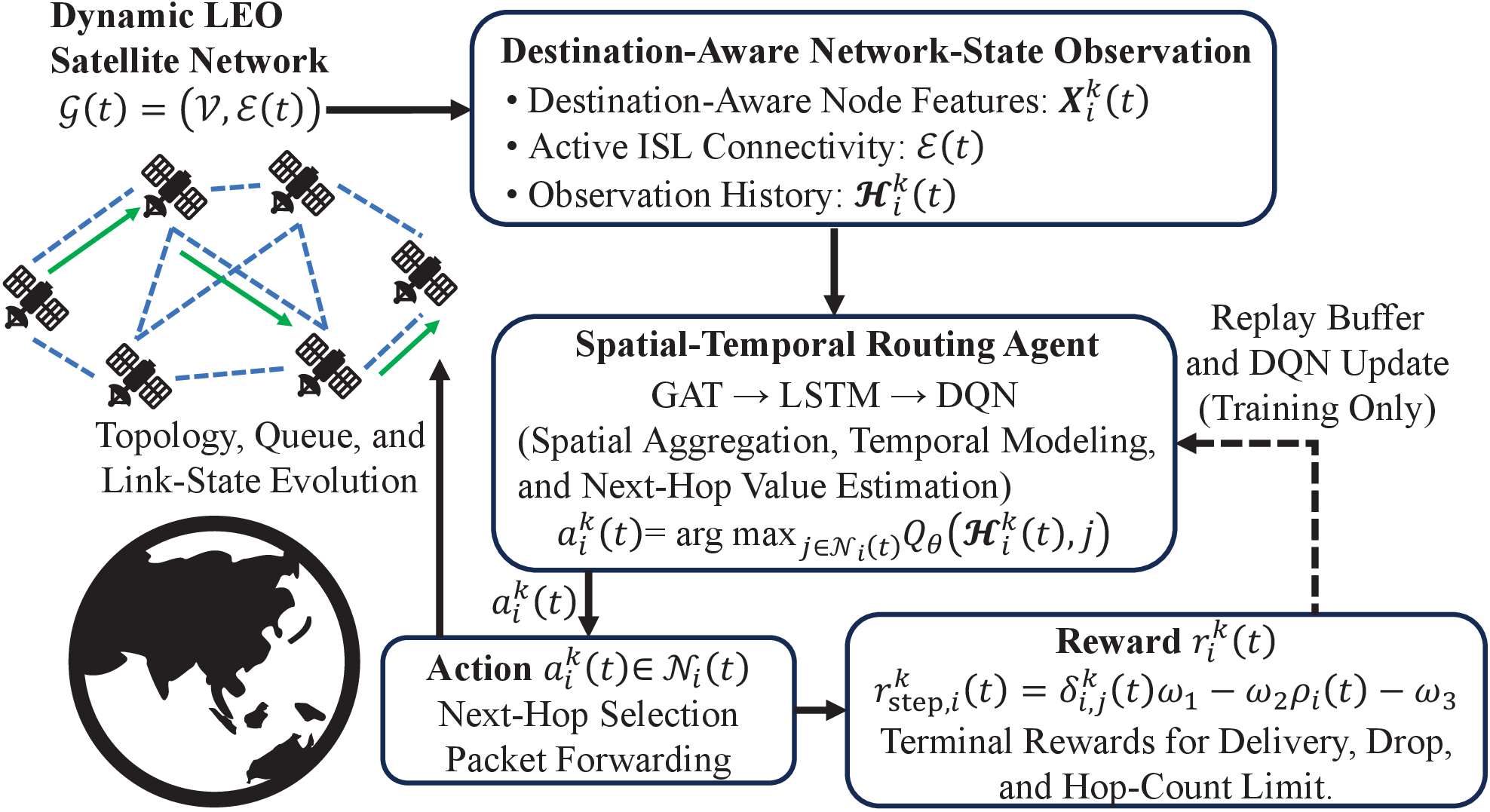}
\caption{Proposed destination-aware spatial-temporal routing framework.}
\label{fig:system_model}
\vspace{-0.1in}
\end{figure}

Fig.~\ref{fig:system_model} illustrates
the proposed GAT--LSTM--DQN framework
for distributed destination-aware routing.

\subsection{Network Model}

We consider a Walker-Delta LEO constellation composed of
$N$ satellites distributed over multiple orbital planes.
Let $t$ denote the discrete time-slot index.
The satellite network is modeled as a time-varying graph
$\mathcal{G}(t)=(\mathcal{V},\mathcal{E}(t))$
~\cite{Xu2022SpatialLocationAidedFDR}, where
$\mathcal{V}=\{1,2,\ldots,N\}$ denotes the satellite set
and $\mathcal{E}(t)\subseteq\mathcal{V}\times\mathcal{V}$
denotes the active ISL set.

Each satellite maintains intra-plane ISLs with its adjacent
satellites. Inter-plane ISLs are established through
deterministic one-to-one matching between eligible satellites
in adjacent orbital planes. Link availability is subject to
line-of-sight (LOS) constraints, while inter-plane connections
are restricted to satellite latitudes within $\pm60^\circ$.
The maximum number of active ISLs per satellite is limited
to four. As satellites move along their orbits, the topology
is periodically updated, resulting in a time-varying
edge set $\mathcal{E}(t)$.

For each satellite $i\in\mathcal{V}$, the currently reachable
one-hop neighbor set is
$\mathcal{N}_i(t)=
\{j\in\mathcal{V}\mid(i,j)\in\mathcal{E}(t)\}$.
For a packet $k$, let $s_k\in\mathcal{V}$ and
$d_k\in\mathcal{V}$ denote its source and destination
satellites, respectively. The routing task is to select
feasible next-hop nodes from $\mathcal{N}_i(t)$ until the
packet reaches $d_k$ or is terminated due to buffer overflow,
unavailable forwarding routes, or the maximum hop-count
constraint.

\vspace{-0.1in}

\subsection{Traffic and Queue Model}

The packet-generation process at satellite $i$ follows a
non-homogeneous Poisson process (NHPP) with time-varying
arrival rate
\begin{equation}
\lambda_i(t)
=
\bar{\lambda}
\left[
1+\beta\sin\left(
\frac{2\pi t}{T}+\frac{2\pi i}{N}
\right)
\right],
\end{equation}
where $\bar{\lambda}$ denotes the average arrival rate,
$\beta$ controls traffic variation, and $T$ denotes
the traffic period in slots ($T=30$ for a 0.6 s
period with 20 ms slots). The aggregate offered load is distributed
among the source satellites. Additional file-transfer
protocol (FTP) sessions are introduced to emulate
bursty traffic conditions.

Each satellite maintains a shared first-in-first-out
(FIFO) queue with capacity $Q_{\max}$. Let $Q_i(t)$
denote the number of queued packets at satellite $i$,
where $0\leq Q_i(t)\leq Q_{\max}$. The packet-level
model tracks each packet's destination, generation
time, and hop count. At each time slot, a satellite
serves at most $\mu$ packets across all outgoing
ISLs, where $\mu$ denotes the per-satellite
service capacity in packets per slot, and each packet can traverse at most one hop.
Packets are dropped upon buffer overflow or
exhaustion of the maximum hop-count constraint.

\vspace{-0.1in}

\subsection{Delay Model}

The end-to-end delay is measured using the packet-level
forwarding process, which accounts for packet waiting
and transmission across successive time slots.
For a successfully delivered packet $k$, the
end-to-end delay is defined as
$D_k^{\mathrm{e2e}}=
(t_k^{\mathrm{del}}-t_k^{\mathrm{gen}}+1)\Delta t$,
where $t_k^{\mathrm{gen}}$ and
$t_k^{\mathrm{del}}$ denote the packet
generation and delivery slot indices,
respectively, and $\Delta t$ is the slot
duration.
The delay is evaluated only for successfully delivered
packets. Propagation, transmission, and processing
delays are considered as diagnostic components and
are not added again to the measured end-to-end delay.

\subsection{POMDP Formulation}

Because the available observations do not fully capture
the underlying traffic dynamics and future network
conditions, the routing problem is formulated as a
POMDP~\cite{Zuo2021DRLLEORouting}, defined
by $(\mathcal{S},\mathcal{A},\mathcal{P},
\mathcal{R},\Omega,\mathcal{O})$,
representing the state space, action space,
transition model, reward function,
observation space, and observation function,
respectively.

\textbf{Global State and Observation:}
The global network state $s(t)\in\mathcal{S}$
includes the instantaneous topology, satellite
positions, queue occupancies, link conditions,
and traffic states of the constellation.

For a packet $k$ forwarded by satellite $i$ toward
destination $d_k$, the observation is a destination-aware
network-wide node-feature matrix
$\bm{X}_i^k(t)\in\mathbb{R}^{N\times F_{\mathrm{in}}}$,
where $F_{\mathrm{in}}=6$. Let $\bm{p}_j(t)\in\mathbb{R}^3$
denote satellite $j$'s position and
$R=R_{\mathrm{E}}+h$ the orbital radius, with
$R_{\mathrm{E}}$ and $h$ denoting Earth's radius
and orbital altitude, respectively. The $j$th row
of $\bm{X}_i^k(t)$ is the feature vector
\begin{equation}
\label{eq:node_feature}
\begin{aligned}
\bm{x}_j^{i,k}(t)
=\Bigg[
&\frac{(\bm{p}_j(t)-\bm{p}_{d_k}(t))^{\mathsf T}}{2R},
\frac{Q_j(t)}{Q_{\max}},\\
&\frac{\|\bm{p}_j(t)-\bm{p}_{d_k}(t)\|_2}{2R},
\mathds{1}\{j=i\}
\Bigg]^{\mathsf T}.
\end{aligned}
\end{equation}
where $\mathds{1}\{\cdot\}$ is the indicator function.
The six components comprise three destination-relative
position coordinates, normalized queue occupancy,
normalized distance to the destination, and an indicator
of the current forwarding satellite. Thus, destination
information is embedded in the node features rather
than supplied as a separate input vector.
The GAT additionally uses the adjacency structure
constructed from $\mathcal{E}(t)$ with self-loops.

To capture temporal network-state variations, the
routing agent maintains an ordered packet-specific
history $\mathcal{H}_i^k(t)$ of the most recent
$T_w$ destination-aware feature matrices. In the
implementation, this history is maintained for each
packet as it traverses the network; the first observed
matrix is repeated to initialize the window. During GAT processing, the adjacency mask
of the evaluated snapshot is applied to
the historical node-feature matrices.

Although the observation incorporates network-wide
satellite positions and queues, forwarding actions are
restricted to currently available one-hop neighbors.
Therefore, next-hop decisions are made independently
at individual satellites rather than by centralized
end-to-end route computation. Acquiring network-wide
state information may nevertheless require signaling.

\textbf{Action:}
When satellite $i$ holds packet $k$, its feasible action set is determined by the currently active ISLs, i.e.,
$\mathcal{A}_i(t)=\mathcal{N}_i(t)$.
The action $a_i^k(t)\in\mathcal{A}_i(t)$ corresponds to selecting one neighboring satellite as the next-hop node.

\noindent\textbf{State Transition:}
The transition probability
$\mathcal{P}\bigl(s(t+1)\mid
s(t),a_i^k(t)\bigr)$
is governed by satellite mobility, ISL activation and deactivation, stochastic packet arrivals, queue evolution, and packet-forwarding decisions. Since these coupled dynamics are difficult to characterize exactly, $\mathcal{P}$ is treated as an unknown environment function, motivating the use of model-free DRL.

\textbf{Reward:}
The instantaneous reward encourages successful delivery, destination progress, and congestion avoidance while penalizing packet drops and unnecessary forwarding. It is defined as
\begin{equation}
r_i^k(t)
=
\begin{cases}
R_{\mathrm{success}},
& \text{if the packet is delivered},\\
-R_{\mathrm{drop}},
& \text{if the packet is dropped},\\
r_{\mathrm{step},i}^{k}(t),
& \text{otherwise}.
\end{cases}
\end{equation}
For nonterminal forwarding decisions, the step
reward is $r_{\mathrm{step},i}^{k}(t)=
\delta_{i,j}^{k}(t)\omega_1-
\omega_2\rho_i(t)-\omega_3$,
where $\rho_i(t)=Q_i(t)/Q_{\max}$
denotes the queue-occupancy ratio of the
current satellite before forwarding.
The destination-progress indicator
$\delta_{i,j}^{k}(t)$ equals $+1$ if
$d_{j,d_k}(t)<d_{i,d_k}(t)$ and $-1$
otherwise, where $j=a_i^k(t)$ and
$d_{u,d_k}(t)$ denotes the distance
from satellite $u$ to destination $d_k$.

Accordingly, $\delta_{i,j}^{k}(t)$ indicates whether the selected next hop moves the packet closer to its destination. In the simulations, we set
$R_{\mathrm{success}}=10$,
$R_{\mathrm{drop}}=10$,
$\omega_1=1$,
$\omega_2=2$, and
 $\omega_3=0.1$.
The coefficients $\omega_1$, $\omega_2$, and
$\omega_3$ weight destination progress, queue
occupancy, and the per-hop penalty, respectively.
 Packets exceeding the maximum hop count are terminated with a reward of $-10$.

The routing policy is trained to maximize the expected discounted cumulative reward with discount factor $\gamma\in(0,1)$.

\section{Proposed Spatial-Temporal Learning-Based Distributed Routing Scheme}
\label{sec:proposed}

The proposed framework integrates a shared
feature encoder, GAT, LSTM, and DQN for
destination-aware next-hop selection.
For packet $k$ at satellite $i$, the model
processes the network-wide observation
history defined in Sec.~\ref{sec:system_model}
to estimate Q-values for all $N$ satellites.
GAT extracts spatial representations using
the active ISL connectivity, while LSTM
captures temporal variations in node features.
An action mask restricts the final decision
to the feasible neighbor set $\mathcal N_i(t)$.
During training, an $\epsilon$-greedy policy
is adopted, whereas deployment uses greedy
next-hop selection without centralized
end-to-end route computation.

\subsection{Spatial Feature Extraction via GAT}

Following the GAT formulation in~\cite{Velickovic2018GAT},
we first apply a shared feature encoder to each
satellite's six-dimensional feature vector. For satellite $j$, let
$\bm x_j^{i,k}(t)$ be the $j$th row of
$\bm X_i^k(t)$. Its encoded representation is
$\bm e_j^{i,k}(t)=\operatorname{ReLU}
(\bm W_{\mathrm e}\bm x_j^{i,k}(t)+\bm b_{\mathrm e})$,
where $\operatorname{ReLU}(\cdot)$ denotes the
rectified linear unit, and
$\bm W_{\mathrm e}\in
\mathbb R^{64\times F_{\mathrm{in}}}$ and
$\bm b_{\mathrm e}\in\mathbb R^{64}$
are trainable.

The GAT has $K=4$ attention heads, each with
$F_h=16$ output features, giving a total of
$F_g=KF_h=64$ features. For head $m$, let
$\widetilde{\bm e}_{j}^{(m),i,k}(t)
=\bm W^{(m)}\bm e_j^{i,k}(t)$,
where $\bm W^{(m)}\in\mathbb R^{F_h\times64}$.
The input graph has self-loops, with
$\overline{\mathcal N}_j(t)=\{j\}\cup\mathcal N_j(t)$.
For $v\in\overline{\mathcal N}_j(t)$, the
attention coefficient of head $m$ is
\begin{equation}
\label{eq:gat_attention}
\alpha_{jv}^{(m),i,k}(t)=
\frac{\exp\!\left(\phi(s_{jv}^{(m),i,k}(t))\right)}
{\sum_{u\in\overline{\mathcal N}_j(t)}
\exp\!\left(\phi(s_{ju}^{(m),i,k}(t))\right)},
\end{equation}
where $\phi(\cdot)$ is LeakyReLU with negative
slope $0.2$ and
$s_{jv}^{(m),i,k}(t)=
(\bm a_{\mathrm s}^{(m)})^{\mathsf T}
\widetilde{\bm e}_{j}^{(m),i,k}(t)+
(\bm a_{\mathrm t}^{(m)})^{\mathsf T}
\widetilde{\bm e}_{v}^{(m),i,k}(t)$.
The trainable vectors
$\bm a_{\mathrm s}^{(m)},\bm a_{\mathrm t}^{(m)}
\in\mathbb R^{F_h}$ score source and target
features, respectively.

The head outputs are concatenated and passed
through an exponential linear unit (ELU):
\begin{equation}
\label{eq:gat_spatial}
\bm z_j^{i,k}(t)=\operatorname{ELU}\!\left(
\mathop{\Vert}_{m=1}^{K}
\sum_{v\in\overline{\mathcal N}_j(t)}
\alpha_{jv}^{(m),i,k}(t)
\widetilde{\bm e}_{v}^{(m),i,k}(t)
\right),
\end{equation}
where $\Vert$ denotes concatenation.
Thus, $\bm z_j^{i,k}(t)\in\mathbb R^{64}$.
Although the implementation computes dense
attention scores for all node pairs, unavailable
ISLs are excluded by the adjacency mask.

\subsection{Temporal Dependency Modeling via LSTM}

To capture temporal network-state variations,
we apply an LSTM~\cite{Hochreiter1997LSTM}
to each satellite's sequence of spatial
representations. For candidate satellite $j$,
the temporally ordered input is
$\bm Z_{i,j}^k(t)=
[\bm z_j^{i,k}(t-T_w+1),\ldots,
\bm z_j^{i,k}(t)]$,
where $T_w=5$. The LSTM uses a 64-dimensional
input and a 128-dimensional hidden state,
producing $\bm h_{i,j}^k(t)\in\mathbb R^{128}$
from the final time step.

Each packet's feature history is initialized
by repeating its first observation.
The adjacency mask of the evaluated snapshot
is applied throughout the input window.
Thus, temporal variations are captured from
historical node features without a separate
traffic-prediction stage.

\subsection{DRL-Based Routing Decision}

The shared Q-head processes the LSTM representation
of each candidate satellite to obtain its Q-value.
Specifically, with trainable matrices and biases,
\begin{equation}
\label{eq:q_head}
Q_\theta(\bm{\mathcal H}_i^k(t),j)
=\bm w_{\mathrm q}^{\mathsf T}
\operatorname{ReLU}(\bm W_{\mathrm q}
\bm h_{i,j}^k(t)+\bm b_{\mathrm q})+b_{\mathrm q},
\end{equation}
where $\bm W_{\mathrm q}\in\mathbb R^{64\times128}$,
$\bm w_{\mathrm q},\bm b_{\mathrm q}\in\mathbb R^{64}$,
and $b_{\mathrm q}\in\mathbb R$ are trainable parameters.
Here, $\bm{\mathcal H}_i^k(t)$ denotes the
ordered collection of the packet's historical
feature matrices, together with the adjacency
mask used for the evaluated snapshot.

The network therefore produces $N$ Q-values
(45 in our simulation), indexed by absolute
satellite IDs. During exploitation, the next hop is
\begin{equation}
\label{eq:next_hop}
a_i^k(t)=\arg\max_{j\in\mathcal N_i(t)}
Q_\theta(\bm{\mathcal H}_i^k(t),j).
\end{equation}
Unavailable neighbors are excluded by action
masking. In greedy tie cases, the smallest
satellite ID is selected. During training, a uniformly sampled feasible
neighbor is chosen with exploration probability
$\epsilon$; otherwise, Eq.~\eqref{eq:next_hop}
is used.

The experience replay buffer $\mathcal D$
stores packet-level transitions containing
the current and next state sequences, chosen
satellite ID, reward, terminal flag, and the
corresponding adjacency and next-action masks.
For a transition whose next forwarding satellite
is $j=a_i^k(t)$, the temporal-difference target is
$y_i^k(t)=r_i^k(t)+\gamma b_i^k(t)$,
where $\theta$ and $\theta^-$ denote the online-
and target-network parameters, respectively,
$\gamma$ is the discount factor, and
\begin{equation}
\label{eq:dqn_bootstrap}
b_i^k(t)=\begin{cases}
\displaystyle\max_{a'\in\mathcal N_j(t+1)}
Q_{\theta^-}(\bm{\mathcal H}_j^k(t+1),a'),
&\text{if valid},\\
0,&\text{otherwise},
\end{cases}
\end{equation}
where ``valid'' means that the transition is
nonterminal and $\mathcal N_j(t+1)\ne\varnothing$.
Thus, bootstrap values are omitted for delivered
or dropped packets and when the next feasible
neighbor set is empty.

The online Q-network minimizes the SmoothL1 loss
\begin{equation}
\label{eq:dqn_loss}
\mathcal L(\theta)=\mathbb E_{\mathcal D}\!\left[
\ell_{\mathrm{SL1}}\!\left(
 y_i^k(t)-Q_\theta(\bm{\mathcal H}_i^k(t),a_i^k(t))
\right)\right],
\end{equation}
where $\ell_{\mathrm{SL1}}$ denotes the
SmoothL1 loss, and $\mathbb E_{\mathcal D}$
denotes mini-batch averaging over replay
samples. The online network is updated
using Adam with gradient clipping, while
the target network is periodically
synchronized.

\subsection{Distributed Routing Mechanism}

The shared policy enables independent
next-hop selection at each satellite.
Acquiring network-wide observations may
introduce signaling overhead, which is
not evaluated in this work.

\section{Simulation Results}
\label{sec:simulation}

\subsection{Simulation Setup}

The proposed scheme is evaluated in a dynamic
Walker-Delta LEO constellation comprising
45 satellites distributed over three orbital
planes. The orbital altitude and inclination
are 570 km and $70^\circ$, respectively.
The network topology is updated every 20 ms
according to satellite mobility and ISL
availability. Each satellite maintains a FIFO
queue with a capacity of 100 packets and
serves at most 36 packets per time slot.

Background packet arrivals follow the NHPP model in
Sec.~\ref{sec:system_model}, with an aggregate offered load of 240 Mbps
across all four scenarios. Four traffic scenarios are considered:
local and cross-region routing under normal
traffic and FTP burst traffic. Each FTP
session contains 2 MB of data, and the packet size is 1500 bytes. The overall packet loss
rate is defined as the ratio of dropped packets
to generated packets. The main network
parameters are summarized in
Table~\ref{tab:network_params}.

The proposed method is compared with three
adapted learning-based routing baselines:
DQN-IR-adapted~\cite{Zuo2021DRLLEORouting},
FDR-MARL-adapted~\cite{Xu2022SpatialLocationAidedFDR},
and GraphPR-adapted~\cite{Ran2025GNNEnhancedMARL}.
These baselines are evaluated in the same
simulation environment using paired traffic
realizations. Five independent training and
testing seed pairs are used for performance
evaluation. The learning-based schemes follow
a common training-transition budget of
approximately $10^7$ transitions.
The proposed model hyperparameters are
summarized in Table~\ref{tab:drl_params}.

The implementation, including the adapted
baselines, simulation configurations, and
evaluation scripts, is publicly available
for reproducibility.\footnote{Source code
and configuration files are available at:
\texttt{github.com/BoneZhou/Spatial-Temporal-LEO-Routing}.}

\begin{table}[t]
\centering
\caption{Simulation Parameters of the LEO Satellite Network}
\label{tab:network_params}
\resizebox{0.65\linewidth}{!}{
\begin{tabular}{lc}
\hline
\textbf{Parameter} & \textbf{Value} \\
\hline
Number of satellites & 45 \\
Number of orbital planes & 3 \\
Orbital altitude & 570 km \\
Orbital inclination & $70^\circ$ \\
Topology update interval & 20 ms \\
Maximum active ISLs per satellite & 4 \\
Inter-plane latitude limit & $\pm60^\circ$ \\
ISL capacity & 300 Mbps \\
Aggregate offered load & 240 Mbps \\
Packet size & 1500 bytes \\
Queue capacity per satellite & 100 packets \\
Service capacity per satellite & 36 packets/slot \\
Maximum hop count & 30 \\
Traffic variation factor & 0.35 \\
Traffic variation period & 0.6 s \\
FTP session size & 2 MB \\
\hline
\end{tabular}
}
\end{table}

\begin{table}[t]
\centering
\caption{Hyperparameters of the Proposed Framework}
\label{tab:drl_params}
\resizebox{0.58\linewidth}{!}{
\begin{tabular}{lc}
\hline
\textbf{Hyperparameter} & \textbf{Value} \\
\hline
Number of GAT attention heads & 4 \\
GAT hidden dimension & 64 \\
LSTM hidden dimension & 128 \\
Training transition budget & $\approx 10^7$ \\
Learning rate & $1\times10^{-4}$ \\
Discount factor $\gamma$ & 0.99 \\
Replay buffer size & 100,000 \\
Batch size & 128 \\
Target update frequency & 200 steps \\
Initial $\epsilon$ & 1.0 \\
Minimum $\epsilon$ & 0.01 \\
$\epsilon$ decay & 0.995 \\
Observation window $T_w$ & 5 \\
Gradient clipping threshold & 10 \\
\hline
\end{tabular}
}
\end{table}

\subsection{Performance Comparison}

\begin{figure*}[t]
\centering
\includegraphics[width=.83\textwidth]
{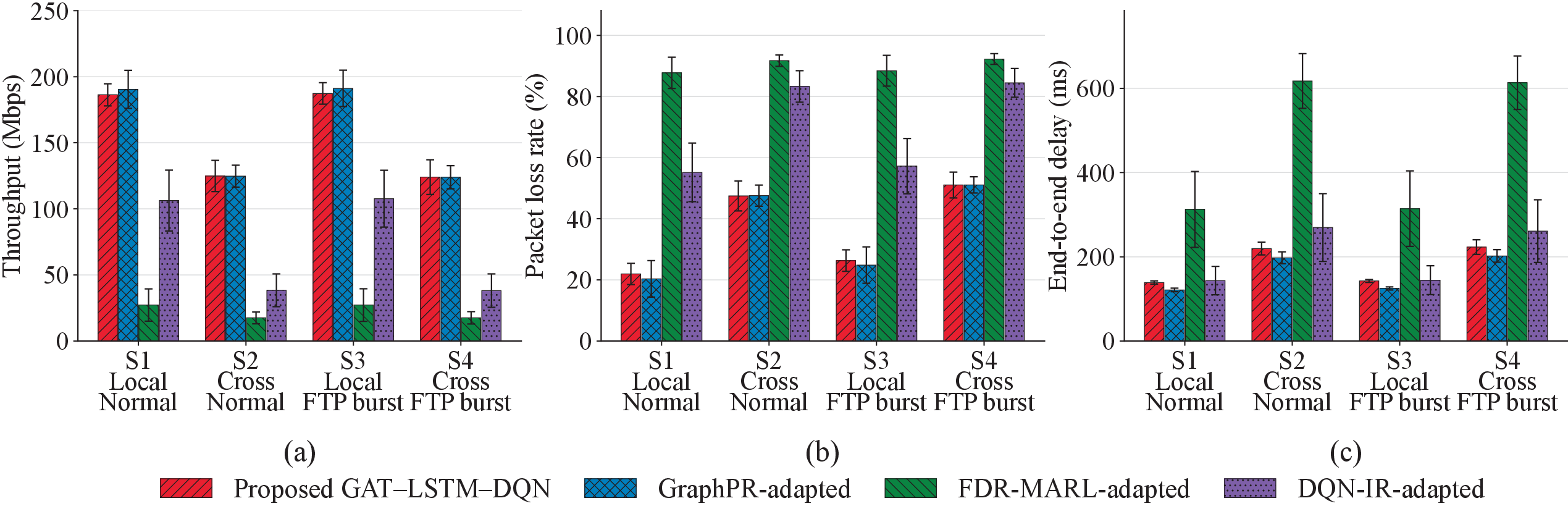}
\caption{Performance comparison under
four traffic scenarios: (a) throughput,
(b) packet loss rate, and (c) end-to-end delay.
Error bars indicate $\pm$1 standard deviation
across five paired seeds.}
\label{fig:performance_comparison}
\end{figure*}

Fig.~\ref{fig:performance_comparison}
compares throughput, packet loss rate,
and end-to-end delay under four traffic
scenarios: S1 (local/normal), S2
(cross-region/normal), S3 (local/burst),
and S4 (cross-region/burst).

As shown in Fig.~\ref{fig:performance_comparison}(a)
and (b), the proposed method achieves
throughput and packet loss performance
comparable to GraphPR across all scenarios.
Under S2, the proposed method achieves
124.85 Mbps throughput and 47.47\% packet
loss, compared with 124.72 Mbps and
47.55\% for GraphPR, respectively.

Fig.~\ref{fig:performance_comparison}(c)
shows that GraphPR achieves lower
end-to-end delay. Under S2, the proposed
method incurs 219.58 ms delay, compared
with 197.55 ms for GraphPR. The corresponding
average queue lengths are 56.19 and
50.85 packets, respectively.
These results indicate consistent and
competitive packet delivery performance
across different traffic scenarios,
although the proposed method incurs
higher delay and queue occupancy than
GraphPR.

\subsection{Ablation Study}

To evaluate the contributions of spatial
and temporal learning, we compare the
proposed GAT--LSTM--DQN framework with
three variants: GAT-DQN, LSTM-DQN, and
conventional DQN. GAT-DQN removes the
temporal module, LSTM-DQN removes the
spatial module, and DQN excludes both.

\begin{figure*}[t]
\centering
\includegraphics[width=.83\textwidth]
{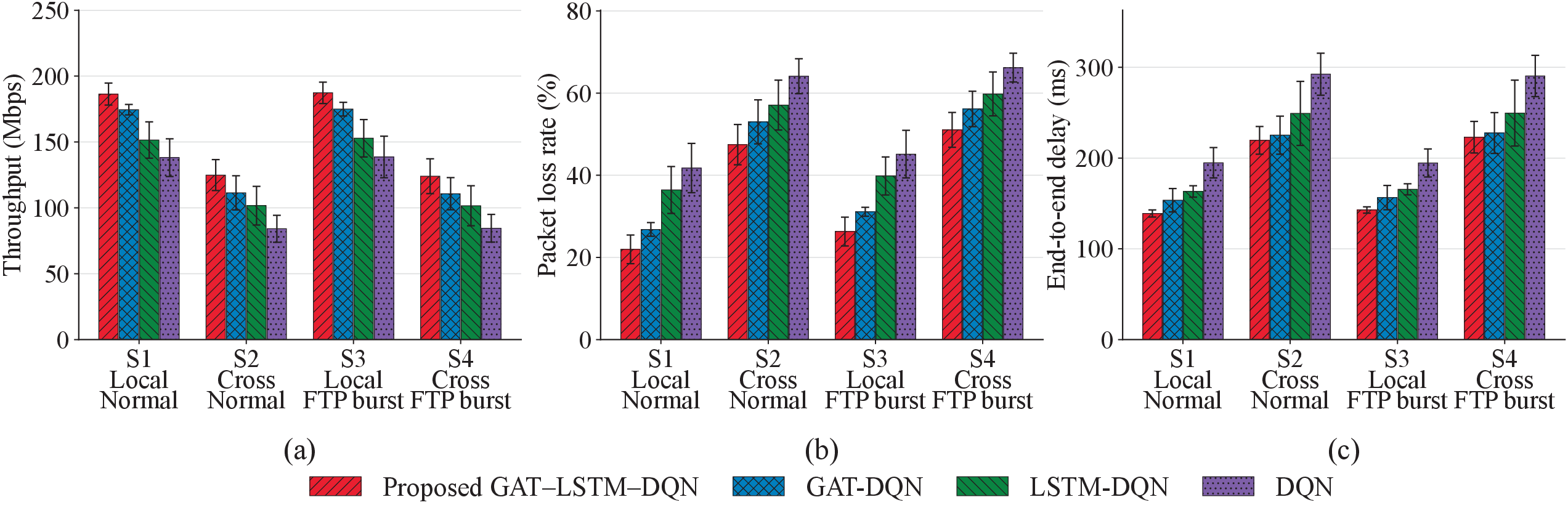}
\caption{Ablation study under four traffic
scenarios: (a) throughput, (b) packet loss
rate, and (c) end-to-end delay.
Error bars indicate $\pm$1 standard deviation
across five paired seeds.}
\label{fig:ablation}
\end{figure*}

Fig.~\ref{fig:ablation} compares the
three performance metrics under S1--S4.
The proposed framework achieves higher
throughput, lower packet loss rates, and
shorter end-to-end delay than the three
variants across all four scenarios.
These results demonstrate the complementary
benefits of spatial and temporal learning
for distributed next-hop selection.

\subsection{Dynamic Traffic Analysis}

\begin{figure}[t]
\centering
\includegraphics[width=.83\columnwidth]
{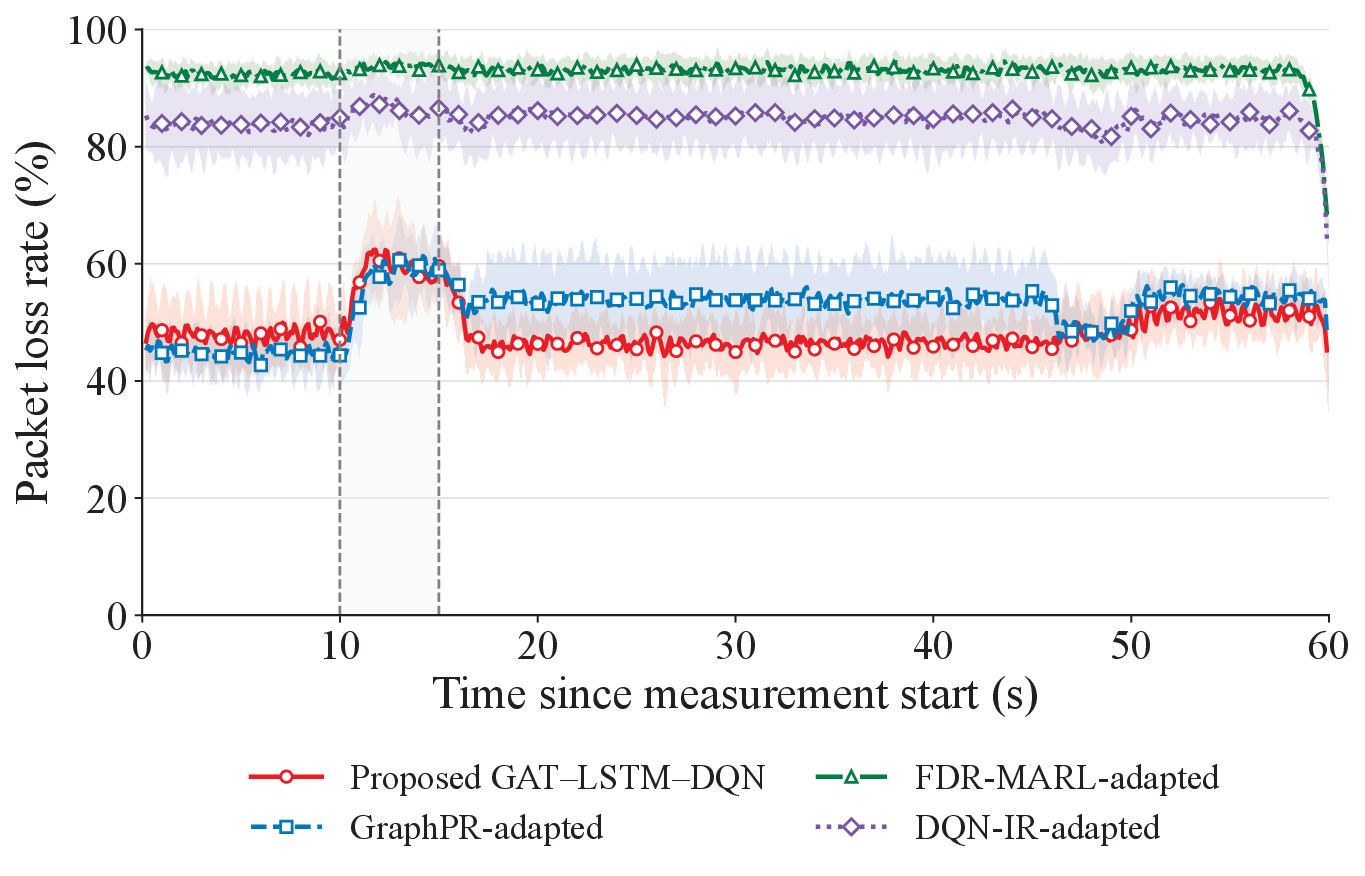}
\caption{Temporal evolution of the 0.2-s rolling
birth-cohort packet loss rate under cross-region
FTP burst traffic between 10 and 15 s.
Solid lines show the mean, and shaded regions
indicate $\pm$1 standard deviation across
five paired seeds.}
\label{fig:dynamic_traffic}
\end{figure}

Fig.~\ref{fig:dynamic_traffic} shows packet loss rates over 60 s
with an FTP burst between 10 and 15 s.
The proposed method and GraphPR exhibit
comparable loss rates during the burst.
Afterward, the proposed method maintains
lower packet loss over most of the
observation period, suggesting improved
post-burst packet delivery reliability.

\subsection{Computational Complexity Analysis}

Let $N$, $F_{\mathrm{in}}$, and $E$ denote
the satellite count, input dimension,
and encoder output dimension, respectively.
The GAT uses $K$ heads with $F_h$
features per head, yielding $D=KF_h$.
Let $T_w$, $H$, and $U$ denote the
observation-window length, LSTM hidden
dimension, and Q-head hidden dimension.
With dense attention, the per-decision
inference complexity is
\begin{equation}
\label{eq:overall_complexity}
\begin{aligned}
C_{\mathrm{total}}=\mathcal{O}\big(
&T_w N F_{\mathrm{in}}E
+T_w KNEF_h\\
&+T_w KN^2F_h
+NT_w(DH+H^2+H)\\
&+N(HU+U)
\big).
\end{aligned}
\end{equation}

In our implementation,
$(F_{\mathrm{in}},E,K,F_h,T_w,H,U)
=(6,64,4,16,5,128,64)$.
Action masking costs $\mathcal{O}(N)$.
Network-wide observation signaling
overhead is excluded.

\section{Conclusion}

This paper proposed a GAT--LSTM--DQN
framework for distributed routing in
dynamic LEO satellite networks.
Simulations demonstrated consistent,
competitive packet delivery performance
across different traffic scenarios,
although GraphPR achieved lower delay
and queue occupancy. Ablation results
confirmed the complementary benefits
of spatial and temporal learning.
Future work will address signaling
overhead and congestion-aware routing.


\begin{thebibliography}{99}

\bibitem{Chen2026HMLC}
G. Chen, T. Liao, S. Meng, and S. Wu,
``Heterogeneous multi-layer constellation for future satellite internet: Framework, optimization, and evolution vision,''
\emph{IEEE Wireless Commun.}, vol. 33, no. 3, pp. 14--22, Jun. 2026.

\bibitem{Dong2023DRL}
F. Dong, J. Song, Y. Zhang, Y. Wang, and T. Huang,
``DRL-based load-balancing routing scheme for 6G space--air--ground integrated networks,''
\emph{Remote Sens.}, vol. 15, no. 11, p. 2801, May 2023.

\bibitem{Xu2022SpatialLocationAidedFDR}
G. Xu, Y. Zhao, Y. Ran, R. Zhao, and J. Luo,
``Spatial location aided fully-distributed dynamic routing for large-scale LEO satellite networks,''
\emph{IEEE Commun. Lett.}, vol. 26, no. 12, pp. 3034--3038, Dec. 2022.

\bibitem{Chen2025TransformerMIXLEO}
X. Chen, Z. Ji, S. Wu, H. Jia, A. Xiao, and C. Jiang,
``A distributed routing algorithm for LEO satellite networks: A multiagent transformer-MIX learning approach,''
\emph{IEEE Internet Things J.}, vol. 12, no. 11, pp. 15748--15763, Jun. 2025.

\bibitem{Kaelbling1996}
L. P. Kaelbling, M. L. Littman, and A. W. Moore,
``Reinforcement learning: A survey,''
\emph{J. Artif. Intell. Res.}, vol. 4, pp. 237--285, May 1996.

\bibitem{Mnih2015DQN}
V. Mnih, K. Kavukcuoglu, D. Silver, A. A. Rusu, J. Veness,
M. G. Bellemare, A. Graves, M. Riedmiller, A. K. Fidjeland,
G. Ostrovski, S. Petersen, C. Beattie, A. Sadik, I. Antonoglou,
H. King, D. Kumaran, D. Wierstra, S. Legg, and D. Hassabis,
``Human-level control through deep reinforcement learning,''
\emph{Nature}, vol. 518, no. 7540, pp. 529--533, Feb. 2015.

\bibitem{Chou2026ICCWkshps}
P.-H. Chou, C. Wang, K.-H. Chen, and W.-C. Hsiao,
``DRL-based beam positioning for LEO satellite constellations with weighted least squares,''
in \emph{Proc. IEEE Int. Conf. Commun. Workshops (ICC Wkshps)},
May 2026, pp. 1--6.

\bibitem{Zuo2021DRLLEORouting}
P. Zuo, C. Wang, Z. Yao, S. Hou, and H. Jiang,
``An intelligent routing algorithm for LEO satellites based on deep reinforcement learning,''
in \emph{Proc. IEEE 94th Veh. Technol. Conf. (VTC-Fall)}, Sep. 2021, pp. 1--5.

\bibitem{Li2025ParetoMARLQueueing}
S. Li, G. Wu, Q. Wu, R. Wang, and H. Zhang,
``Efficient packet routing for large-scale LEO satellite networks: A pareto-optimal MARL approach with queueing theory,''
\emph{IEEE Internet Things J.}, vol. 12, no. 22, pp. 46675--46691, Nov. 2025.

\bibitem{Zhang2025GRLR}
S. Zhang, A. Liu, C. Han, X. Xu, X. Liang, K. An, and Y. Zhang,
``GRLR: Routing with graph neural network and reinforcement learning for mega LEO satellite constellations,''
\emph{IEEE Trans. Veh. Technol.}, vol. 74, no. 2, pp. 3225--3237, Feb. 2025.

\bibitem{Ran2025GNNEnhancedMARL}
Y. Ran, Y. Ding, S. Chen, J. Lei, and J. Luo,
``Fully-distributed dynamic packet routing for LEO satellite networks: A GNN-enhanced multi-agent reinforcement learning approach,''
\emph{IEEE Trans. Veh. Technol.}, vol. 74, no. 3, pp. 5229--5234, Mar. 2025.

\bibitem{Rao2025DeepGraphAttentionIES}
Z. Rao, Z. Zhu, D. Niyato, Y. Yao, Y. Xu, and Y. Cheng,
``Dynamic LEO satellite routing approach based on deep graph attention and incremental evolutionary reinforcement learning,''
\emph{IEEE Internet Things J.}, vol. 12, no. 23, pp. 50126--50142, Dec. 2025.

\bibitem{Xiang2025TemporalGraphConvRouting}
J. Xiang, X. He, Y. Zhao, Z. Xie, and X. Liang,
``Distributed dynamic routing for LEO satellite networks with temporal graph convolutions and imitation acceleration,''
\emph{IEEE Commun. Lett.}, vol. 29, no. 11, pp. 2521--2525, Nov. 2025.

\bibitem{Ju2025SpatioTemporalTrafficPrediction}
Y. Ju, J. Song, W. Li, Y. Zhang, C. He, F. Dong, and C. Chen,
``Dynamic load-balancing routing strategy for LEO satellite networks based on spatio-temporal traffic prediction,''
\emph{IEEE Trans. Aerosp. Electron. Syst.}, vol. 61, no. 5, pp. 11954--11970, Oct. 2025.

\bibitem{Wang2025STStatePredictionRouting}
Y. Wang, Z. Zhu, K. Wu, Y. Hou, H. He, and J. Yang,
``Spatio-temporal correlated network state prediction and dynamic routing for satellite networks,''
in \emph{Proc. IEEE Wireless Commun. Netw. Conf. (WCNC)}, Mar. 2025, pp. 1--7.

\bibitem{Li2026DelayAwareLEOIoT}
P. Li, L. Chen, J. Wang, P. Xin, J. Luo, P. Pan, and C. Jiang,
``Delay-aware routing optimization for LEO-IoT relying on traffic prediction,''
\emph{IEEE Internet Things J.}, vol. 13, no. 2, pp. 3156--3173, Jan. 2026.

\bibitem{Velickovic2018GAT}
P. Veli\v{c}kovi\'c, G. Cucurull, A. Casanova, A. Romero,
P. Li\`o, and Y. Bengio,
``Graph attention networks,''
in \emph{Proc. Int. Conf. Learn. Represent. (ICLR)}, Apr. 2018.

\bibitem{Hochreiter1997LSTM}
S. Hochreiter and J. Schmidhuber,
``Long short-term memory,''
\emph{Neural Comput.}, vol. 9, no. 8, pp. 1735--1780, Nov. 1997.

\end{thebibliography}
\end{document}